\documentclass[fleqn,usenatbib]{mnras}

\usepackage{newtxtext,newtxmath}

\usepackage[T1]{fontenc}

\DeclareRobustCommand{\VAN}[3]{#2}
\let\VANthebibliography\thebibliography
\def\thebibliography{\DeclareRobustCommand{\VAN}[3]{##3}\VANthebibliography}

\usepackage{graphicx}	
\usepackage{amsmath}	
\usepackage[T2A]{fontenc}
\usepackage[utf8]{inputenc}
\usepackage[english]{babel}
\usepackage[normalem]{ulem}  
\usepackage{xcolor}
\graphicspath{{./}{LEGACY/}}

\newcommand{\noopsort}[1]{}

\title[The heliosheath flow]{Effects of charge exchange on plasma flow in the heliosheath and astrosheathes}

\author[S.D. Korolkov, V.V. Izmodenov]{
S. D. Korolkov,${}^1{}^2{}^3$\thanks{E-mail: korolkov.msu@mail.ru}
V. V. Izmodenov,${}^1{}^2{}^3$\thanks{E-mail: izmod@cosmos.ru}
\\
$^{1}$Space Research Institute (IKI) of Russian Academy of Sciences, Moscow, Russia\\
$^{2}$Lomonosov Moscow State University, Moscow center for fundamental and applied mathematics, Moscow, Russia\\
$^{3}$HSE University, 20 Myasnitskaya Ulitsa, Moscow 101000, Russia
}

\date{Accepted XXX. Received YYY; in original form ZZZ}

\pubyear{2023}

\begin{document}
\label{firstpage}
\pagerange{\pageref{firstpage}--\pageref{lastpage}}
\maketitle

\begin{abstract}
Shock boundary layers are regions bounded by a shock wave on the one side and tangential discontinuity on the other side. These boundary layers are commonly observed in astrophysics. For example, they exist in the regions of the interaction of the stellar winds with the surrounding interstellar medium. Additionally, the shock layers are often penetrated by the flows of interstellar atoms, as, for instance, in the astrospheres of the stars embedded by the partially ionized interstellar clouds. This paper presents a simple toy model of a shock layer that aims to qualitatively describe the influence of charge exchange with interstellar hydrogen atoms on the plasma flow in astrospheric shock layers. To clearly explore this effect, magnetic fields are neglected, and the geometry is kept as simple as possible. The model explains why the cooling of plasma due to charge exchange in the inner heliosheath leads to an increase in plasma density in front of the heliopause. It also demonstrates that the source of momentum causes changes in the pressure profile within the shock layer. The paper also discusses the decrease in plasma density near the astropause in the outer shock layer in the case of layer heating, which is particularly relevant in light of the Voyager measurements in the heliospheric shock layer.
\end{abstract}

\begin{keywords}
stars: winds -- ISM: atoms --
                Sun: evolution, heliosphere --
                Methods: numerical
\end{keywords}



\section{Introduction}

A shock layer (SL) is a region of gas bounded on one side by a shock wave and on the other side by a tangential discontinuity. In this paper, we focus on large-scale shock layers, which are fairly common in astrophysics. They can be found in problems on the explosion of supernovae, the interaction of stellar winds with each other in systems of binary stars, and the interaction of stellar winds with comets, planetary magnetospheres, or interstellar media. At the same time, the plasma flows of the shock layers are often modified by physical processes such as thermal conduction, radiative losses, charge exchange with neutral atoms, etc. In this paper, we investigate the effect of charge exchange with the hydrogen atoms on the shock layers of the astrospheres.  

\begin{figure}
	\begin{minipage}[h!]{1.00\linewidth}
		\center{\includegraphics[width=1.0\linewidth]{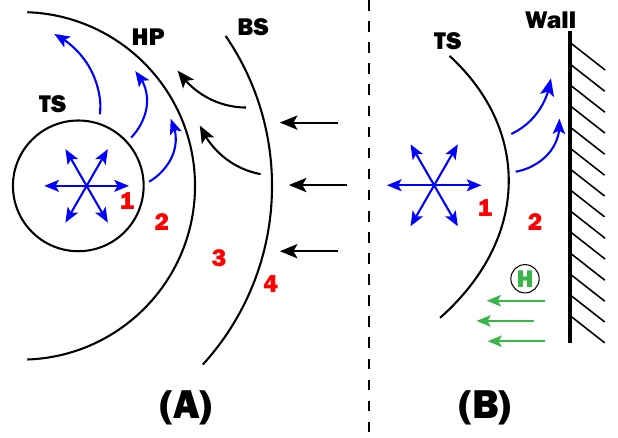} }
	\end{minipage}
	\caption{(A) schematic picture of the main discontinuity surfaces of the astrosphere and numbering of areas. (B) schematic picture of the single shock layer (2) in the problem of the interaction of a source with a planar contact discontinuity/wall}
	\label{fig1}
\end{figure}

The problem of the interaction of the supersonic solar wind with the supersonic flow of the local interstellar medium (LISM) was first formulated and solved in the thin layer approximation by \cite{baranov70}. Both plasma of the solar wind and the interstellar medium were assumed to be fully ionized ideal gases. The structure of such an interaction consists of three main discontinuities (see Figure~\ref{fig1}, A): two shock waves (TS and BS) and a tangential discontinuity surface (HP) called the heliopause. The area filled with the solar wind is called the heliosphere. We conditionally divide the interaction region into four zones (see Figure~\ref{fig1}, A): (1) supersonic stellar wind region, (2) inner SL, (3) outer SL, and (4) interstellar medium. 


Similar structures are observed in other stars \citep{Buren1988, Noriega, Decin, Kobul_2016} as well as obtained in simulations of various astrospheres (see, for example, \cite{2020Herbst, 2022Herbst, 2022Light}).  The region filled with stellar wind is called the astrosphere, and the tangential discontinuity is called the astropause. 


In the vicinity of a star, the Local Interstellar Medium (LISM) may be not fully ionized. The neutral component, primarily composed of atomic hydrogen, interacts with protons primary by charge exchange. This interaction involves the exchange of momentum and energy between the charged and neutral components. In the case of the heliosphere the effect of charge exchange is significant. As a result, the main discontinuities (TS, HP, BS) move closer to the Sun by approximately 30-50\% due to interactions with interstellar neutrals (\cite{Baranov_Malama_1993}, see, also, Figure 2 in  \cite{Izmodenov2000}). Furthermore, charge exchange also modifies the plasma flows within the inner and outer shock layers. For example, measurements from the Voyager spacecraft indicate an increase in proton number density near the HP in the inner shock layer (the region 2 in Figure~\ref{fig1}, A) and a sharp decrease in number density near the HP in the outer shock layer (the region 3 in Figure~\ref{fig1}, A) \citep{Richardson2022}.


The impact of charge exchange on plasma flows can vary across different astrospheres. This phenomenon was investigated in studies conducted by \cite{2005Wood, 2021Wood} specifically for M dwarf star astrospheres. The efficiency of charge exchange on plasma flow is determined by the number density of neutrals in the LISM and the size of the astrosphere.

The objective of this paper is to qualitatively investigate the impact of charge exchange on plasma flows in the shock layers of various astrospheres.
To achieve this, we employ a simple model of a single shock layer (see Figure~\ref{fig1}, B), where the shock layer is depicted as region 2. To qualitatively analyze the structure of the shock layer and its response to different sources of momentum and energy caused by charge exchange, we conduct a parametric study in a dimensionless formulation.


The structure of the work is as follows: in Section~\ref{M}, we describe the model and formulate the problem in dimensionless form; in Section~\ref{R}, we show the results and discuss them; in Section~\ref{relev}, we discuss the application of the results to the heliosphere and some astrospheres; in Section~\ref{C}, we summarize the results; and in Appendix~\ref{A1}, we present additional results of modelling and explain the reasons for the formation of the resulting structure of the shock layer.

\section{Model}
\label{M}

\begin{table*}
\caption{Sets of dimensionless parameters of hydrogen in SL used in the calculation}
\label{tab1}
\begin{tabular}{lccccccc}
\hline
Name& ${\mathrm{Kn}}_0$ (range) & $\hat{n}_\mathrm{H_i}$ & $\hat{V}_\mathrm{H}$& $\hat{T}_\mathrm{H}$ & $Q_2$ & $Q_{3,T}$ & atom populations\\
\hline
\hline
Set 1 & 0.87 - 2.93 & 1 & -0.05 & 0.0015 & negative & cooling & 3 and 4\\
\hline
Set 2 & 74 - 600 & 1 & -2.75 & 11 & negative (weak) & heating & 2\\
\hline
  &  & 0.0014 & -12 & 3.4 &  &  & 1\\
Set 3  & 7.2 - 24 & 0.012 & -2.75 & 11 & positive & heating & 2\\
  & & 0.987 & 1 & 0.18 &  &  & 4\\
\hline
\end{tabular}
\end{table*}

To explore the effects of the charge exchange in the shock layers we consider a shock layer that appears in the interaction of the flow of the supersonic spherically symmetric stellar wind, that is the point source of the fully ionized hydrogen plasma, with a flat wall located at the distance $L$ from the star (see Figure~\ref{fig1}, B).  The flat wall mimics the tangential discontinuity (i.e. heliopause or astropause).
The shock appears to decelerate the supersonic flow before it reaches the wall. This shock mimics the termination shock for astrospheres, so we call it the termination shock as well. 


We work in the rest frame connected with the star. The considered problem is axisymmetric. The axis of symmetry we denote as the X-axis that is directed from the star toward the wall. The Y-axis is an axis in a cylindrical coordinate system, indicating the distance to the X-axis. 

Plasma is considered an ideal mono-atomic gas with a constant heat capacity ($\gamma = \frac{5}{3}$) and is described by Euler equations. A single fluid approach is used with the equation of state: $p = 2n_p k_bT$. 

In addition to the plasma component, the neutral component consisting of atomic hydrogen flows through the shock layer. The hydrogen interacts with protons by charge exchange. The momentum and energy are transferred to/from the plasma component due to this interaction. Since the goal of the paper is to explore the effects of charge exchange on the plasma flow, the effects on the neutral component are neglected for simplicity. Therefore, we  assume  velocity $\vec{V}_{\mathrm{H}}$, number density $n_{\mathrm{H}}$ and temperature $T_{\mathrm{H}}$ to be constant.
The velocity of hydrogen is always assumed to be parallel to the X-axis, but the X-component of the velocity may be both positive or negative.


The system of governing equations for plasma is as follows:

\begin{eqnarray}
	\begin{cases}
		\dfrac{\partial \rho}{\partial t} + \text{div}( \rho\vec{V}) = 0,\\[3mm]
		\dfrac{\partial (\rho \vec{V})}{\partial t} + \text{div}( \rho\vec{V} \vec{V} + p\hat{I}) = \vec{Q}_2,\\[3mm]
		\dfrac{\partial E}{\partial t} + \text{div}( (E+p)\vec{V}) = \vec{Q_2} \cdot \vec{V} + Q_{3, T} ,\\[3mm]
	\end{cases}
	\label{sys1}
\end{eqnarray}
where $E = \frac{p}{\gamma-1}+ \frac{\rho V^2}{2}$ - the energy density.

We have especially divided the energy source in the third equation into the thermal component ($Q_{3,T}$) and the component that the momentum source contributes to the total energy ($\vec{Q_2} \cdot \vec{V}$). The expressions for source terms are written in the form of \cite{McNutt}:
\begin{eqnarray}
	\vec{Q}_2 & = & \nu_{\mathrm{H}}\cdot \left( \vec{V}_{\mathrm{H}} - \vec{V} \right)\label{yr2} , \\ 
	Q_{3,T} & = &  \nu_{\mathrm{H}}\cdot \left[  \dfrac{|\vec{V}_{\mathrm{H}} - \vec{V}|^2}{2}  + \dfrac{U^*_{\mathrm{H}}}{U^*_{\mathrm{M},\mathrm{H}}} (c^2_{\mathrm{H}} - c^2) \right], \label{yr3} 
\end{eqnarray}

\begin{eqnarray}
	\nu_{\mathrm{H}} & = & \frac{1}{m_\mathrm{p}} \rho \rho_{\mathrm{H}}  U^*_{\mathrm{M},\mathrm{H}} \sigma_{\mathrm{ex}}(U^*_{\mathrm{M},\mathrm{H}}),\\
	c & = & \sqrt{\dfrac{2k_\mathrm{B} T}{m_\mathrm{p}}},\ \ \ c_\mathrm{H} = \sqrt{\dfrac{2k_\mathrm{B} T_\mathrm{H}}{m_\mathrm{p}}}, \label{yr5}
\end{eqnarray}
where $V,\ \rho,\ T$ are the velocity, density, and temperature of the plasma, $V_\mathrm{H},\ \rho_\mathrm{H},\ T_\mathrm{H}$ are the velocity, density, and temperature of the hydrogen, $m_\mathrm{p}, k_\mathrm{B}$ are proton mass and Boltzmann constant.

$\sigma_{\mathrm{ex}} (U)$ - charge exchange cross-section, depending on the speed.\\[1mm]
$\sigma_{\mathrm{ex}} (U) = (a_1 - a_2\cdot \mathrm{ln}(U))^2  $ in cm$^2$, where U in cm/s,  $a_1 = 1.64 \cdot 10^{-7}, a_2 = 6.95 \cdot 10^{-9} $ \citep{Lindsay2005}\\

The expressions for average speeds in the source term are as follows:
\begin{eqnarray}
	U^*_{\mathrm{H}} & = & \sqrt{|\vec{V}_{\mathrm{H}} - \vec{V}|^2 + \dfrac{4}{\pi}(c^2_{\mathrm{H}} + c^2)} \label{yr6}\\[3mm]
	U^*_{\mathrm{M},\mathrm{H}} & = & \sqrt{|\vec{V}_{\mathrm{H}} - \vec{V}|^2 + \dfrac{64}{9\pi}(c^2_{\mathrm{H}} + c^2)}  \label{yr7}
\end{eqnarray}

It is important to underline that the effects of charge exchange are considered exclusively inside the SL (region 2 in Figure~\ref{fig1}, B), so the source terms are zero in the region of the supersonic wind. This assumption is an element of our 'toy' model to isolate the effects of charge exchange in the SL from others.

The supersonic flow of the stellar wind is determined by two parameters, which are the mass flow rate ($\dot{M} = 4\pi \rho V R^2$) and terminal velocity ($V_0$).  Since the problem is solved numerically, we have to put inner boundary conditions at a certain distance $R_{\mathrm{in}}$ from the star. In the performed calculations, we assumed $R_{\mathrm{in}} = 0.07 L$, where $L$ is the distance from the star to the wall, as it was defined before. The boundary conditions at ${R}_{\mathrm{in}}$ are the following:
$$ V_\mathrm{in} = V_0,\ \ \rho_\mathrm{in} = \dfrac{\dot{M}}{4\pi V_0 R^2_{\mathrm{in}}}, \ \ p_\mathrm{in}= \dfrac{\rho_\mathrm{in} V_\mathrm{in}^2}{\gamma M_{\mathrm{in}}^2},$$ 
where $M_{\mathrm{in}}$ is the Mach number  at  $\hat{R}_{\mathrm{in}}$. We performed calculations with $M_{\mathrm{in}} = 7$. In this case, the thermal pressure $p_\mathrm{in}$ is more than 80 times less than the dynamic pressure $\rho_\mathrm{in} V_\mathrm{in}^2$, so this parameter is not essential.

\subsection{Dimensionless formulation and parameters}

\label{Hydrogen}

The formulated above problem depends on seven parameters: $L,\ \dot{M},\ V_0,\ n_\mathrm{H},\ V_\mathrm{H},\  T_\mathrm{H}$, and $\sigma_{\mathrm{ex},0}= \sigma_{\mathrm{ex}}(V_0)$. The latter parameter is the charge exchange cross-section corresponding to the velocity $V_0$. The value of this parameter is known from the expression for the cross-section given above (see Section~\ref{M}).

We can reformulate the problem in dimensionless form. Let us relate all distances to $L$, all velocities to $V_0$, all densities to $\dot{M}/ (V_0 L^2)$. Dimensionless boundary conditions will now be written as follows:

$$ \hat{V}_\mathrm{in} = 1,\ \ \hat{\rho}_\mathrm{in} = \dfrac{1}{4\pi \hat{R}^2_{\mathrm{in}}}, \ \ \hat{p}_\mathrm{in}= \dfrac{1}{4\pi \gamma \hat{R}^2_{\mathrm{in}}  M_\mathrm{in}^2}.$$

The system of equations in dimensionless form is as follows:

\begin{eqnarray}
	\begin{cases}
 \label{sys2}
		\dfrac{\partial \hat{\rho}}{\partial \hat{t}} + \text{div}( \hat{\rho}\vec{\hat{V}}) = 0,\\[3mm]
		\dfrac{\partial (\hat{\rho} \vec{\hat{V}})}{\partial \hat{t}} + \text{div}( \hat{\rho}\vec{\hat{V}} \vec{\hat{V}} + \hat{p}\hat{I}) = \dfrac{1}{\mathrm{Kn}_0} \vec{\hat{Q}}_2,\\[3mm]
		\dfrac{\partial \hat{E}}{\partial \hat{t}} + \text{div}( (\hat{E}+\hat{p})\vec{\hat{V}}) = \dfrac{1}{\mathrm{Kn}_0} \left( \vec{\hat{Q_2}} \cdot \vec{\hat{V}} + \hat{Q}_{3, T} \right),\\[3mm]
        \vec{\hat{Q}}_2  = \hat{\nu}_{\mathrm{H}}\cdot \left( \vec{\hat{V}}_{\mathrm{H}} - \vec{\hat{V}} \right) , \\[3mm]
    	\hat{Q}_{3,T} =   \hat{\nu}_{\mathrm{H}}\cdot \left[  \dfrac{|\vec{\hat{V}}_{\mathrm{H}} - \vec{\hat{V}}|^2}{2}  + \dfrac{\hat{U}^*_{\mathrm{H}}}{\hat{U}^*_{\mathrm{M},\mathrm{H}}} (\hat{c}^2_{\mathrm{H}} - \hat{c}^2) \right],\\[3mm]
     \hat{\nu}_{\mathrm{H}}  =   \hat{\rho}  \hat{U}^*_{\mathrm{M},\mathrm{H}} \hat{\sigma}_{\mathrm{ex}}(\hat{U}^*_{\mathrm{M},\mathrm{H}}),\ \  \hat{\sigma}_{\mathrm{ex}}(\hat{U}) = \left(1 - \hat{a}_2 \cdot \mathrm{ln}(\hat{U}) \right)^2,\\[3mm]
	\hat{c}  =  \sqrt{2 \hat{T}},\ \ \ \hat{c}_\mathrm{H} = \sqrt{2 \hat{T}_\mathrm{H}},\ \ \ \hat{a}_2 = \dfrac{a_2}{\sqrt{\sigma_{\mathrm{ex}}(V_0)}}, \label{sys8}
	\end{cases}
\end{eqnarray}

Here $\mathrm{Kn_0} = \dfrac{l_0}{L}$ is the Knudsen number that is the ratio of $l_0 = 1/(n_\mathrm{H}\sigma_{\mathrm{ex}}(V_0))$, the mean free path of protons concerning charge exchange, to the characteristic size of the problem L.

The Equations~(\ref{yr6}), (\ref{yr7}) in the dimensionless form remain unchanged.

As a result, three dimensionless parameters of the problem are obtained:
$$\mathrm{Kn}_0 = \dfrac{1}{n_\mathrm{H}\cdot L \cdot \sigma_{\mathrm{ex}}(V_0)},\  \hat{V}_\mathrm{H} = \dfrac{V_\mathrm{H}}{V_0}, \ \hat{T}_\mathrm{H} = \dfrac{2 k_B T_\mathrm{H}}{m_p V_0^2},$$

The source terms in the right part of System~(\ref{sys2}) are inversely proportional to the Knudsen number ($\mathrm{Kn}_0$). It means that they are directly proportional to the number density of atomic hydrogen and the characteristic size, $L$. Therefore, the larger the hydrogen number density and/or $L$, the larger the source terms, and the stronger the effective coupling between the proton and neutral components.
Other dimensionless parameters, $\hat{V}_\mathrm{H}$ and $\hat{T}_\mathrm{H}$ determine the sign of the source terms. If $\hat{V}_\mathrm{H}$ is negative (or positive but smaller than $\hat{V}$) then the source term into the momentum equation is negative. $\hat{T}_\mathrm{H}$ determines the sign of $\hat{Q}_{3, T}$ that influences the plasma internal energy and temperature as it follows from the internal energy balance equation:

$$
\rho \dfrac{dh}{dt} = \dfrac{dp}{dt} + Q_{3,T},\ \ h = c_p T,
$$
where $h$ is enthalpy, $c_p$  is the specific heat capacity at constant pressure.

$Q_{3, T}$ is negative when the thermal velocity of plasma is large enough that the third term in Equation~(\ref{yr3}) overcomes the first two. Positive $Q_{3, T}$ appears when the temperature of the interstellar atoms is large enough.


Since the goal of the paper is to explore the effects of atoms on the plasma flow in the shock layer, we perform a parametric study by varying the three dimensionless parameters. Our strategy is the following, first, we chose the values of $\hat{V}_\mathrm{H}$ and $\hat{T}_\mathrm{H}$, and then explore how the plasma flow changes by varying of the Knudsen number $Kn_0$.

With this strategy, we choose Set 1 as  $\hat{V}_\mathrm{H} = 0.05$ and  $\hat{T}_\mathrm{H} =0.0015$ (see Table~\ref{tab1}). For Set 1 both the momentum and thermal sources are negative. Therefore, it is expected that plasma in the shock layer decelerates and cools due to charge exchange.

For Set 2, $Q_{3, T} > 0$ and $Q_{2} < 0$, which indicates heating and deceleration of the plasma in the shock layer. However, the momentum source is relatively weak, making this set of neutrals mainly focused on demonstrating the impact of heating on the shock layer structure.



For Set 3, $Q_{3, T} > 0$, and $Q_{2} > 0$, which means heating and acceleration of the plasma in the shock layer.
Set 3, in fact, includes sources of three populations of atomic hydrogen having different parameters. The sources are additive, so $\vec{Q}_2 = \sum \vec{Q}_{2,i}$, $Q_{3,T} = \sum Q_{3,T,i}$, where $i$ is number of population. This case requires some simple corrections in  System~(\ref{sys8}). To get this dimensionless system of equations, we assume $n_H = \sum n_{\mathrm{H_i}}$. Then, the source terms of each population should be multiplied by the relative number density of each population $\hat{n}_{\mathrm{H_i}} = n_{\mathrm{H_i}} /\left(  \sum n_{\mathrm{H_i}}\right)$.

The selection of Knudsen number values for each set was done as follows. The upper limit was chosen such that the influence of charge exchange could be observed, but it was relatively weak, making the solution close to gas-dynamic. The lower limit was chosen to ensure that the effect of charge exchange was significant but not catastrophic, resulting in a 3-4 times change in temperature in the layer. Such a choice was made because the greater the impact of hydrogen on the plasma, the greater the reverse impact of plasma on the hydrogen, making the approximation of constant hydrogen parameters incorrect, and it would be necessary to consider a self-consistent approach.

The choice of neutral parameters is in Table~\ref{tab1}, as well as the extent to which the range of Knudsen numbers applies to real astrospheres/heliospheres, is explained in Section~\ref{relev}.

\section{Results and discussion}
\label{R}

\begin{figure*}
	\begin{minipage}[h!]{1.00\linewidth}
		\center{\includegraphics[width=1.0\linewidth]{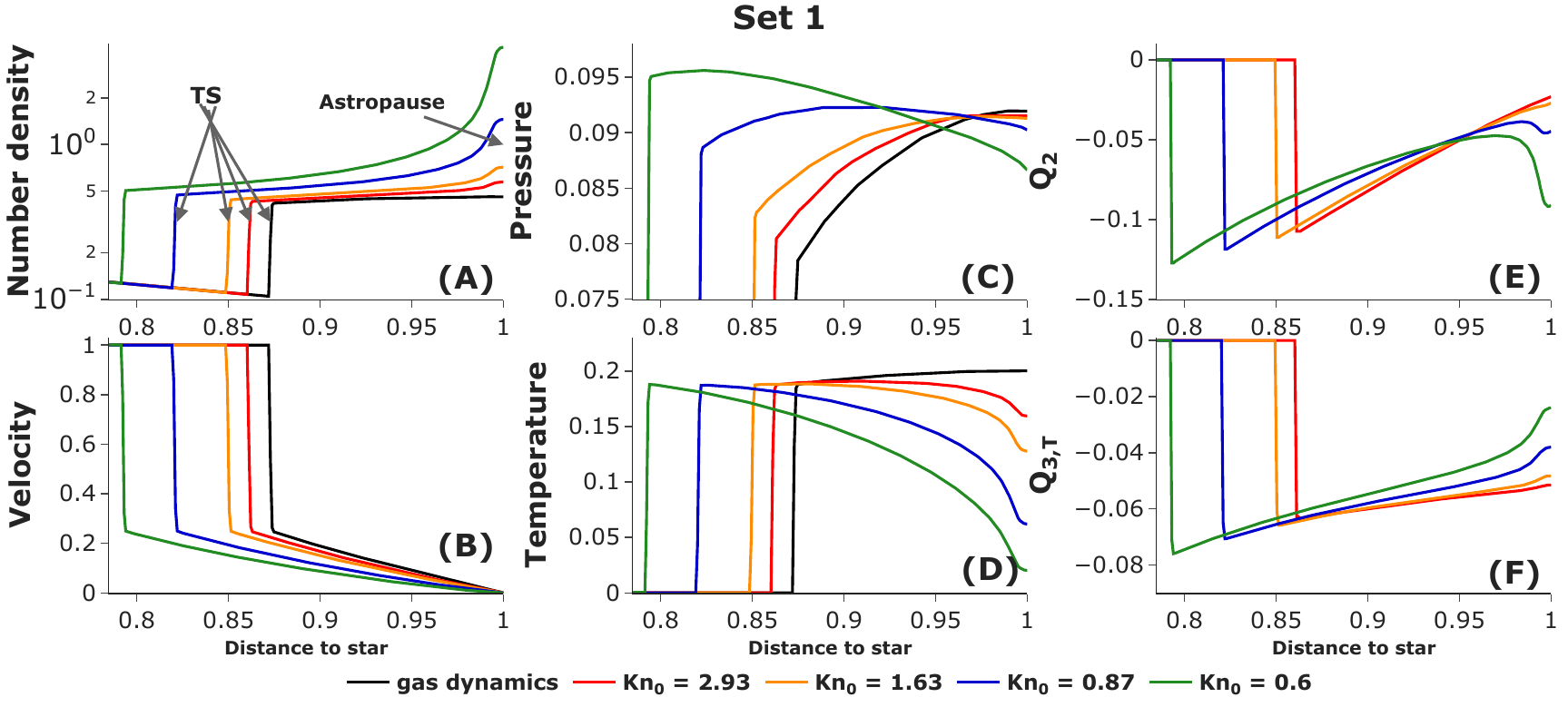} }
	\end{minipage}
	\caption{Distribution of dimensionless plasma parameters in the shock layer on the axis of symmetry (X-axis) for various values of the Knudsen number (including the gas dynamics case) for the distribution of hydrogen in the shock layer given in  Table~\ref{tab1} (Set 1) 
 }
	\label{fig2}
\end{figure*}

In this section, we present results obtained in calculations for different sets of parameters. The summary of the model parameters is given in Table~\ref{tab1}. All plots below are presented for dimensionless parameters.

\subsection{Set 1 results}
\label{s1r}

Figure~\ref{fig2} shows the distribution of plasma parameters in the shock layer along the axis of symmetry (X-axis) for the first set of hydrogen parameters (Set 1, see Table~\ref{tab1}). Different colors of curves correspond to calculations with different Knudsen numbers $\mathrm{Kn}_0$. Black curves correspond to the limiting case of the absence of hydrogen atoms ($\mathrm{Kn}_0 \rightarrow \infty$). In addition to the number density (panel A), velocity (panel B), pressure (panel C), and temperature (panel D) we also plot momentum $\hat{Q}_2$ (panel E) and thermal $\hat{Q}_{3, T}$ (panel F) sources in the right parts of the equations of motion and energy of the System~(\ref{sys2}). For parameters of Set 1, both sources are negative throughout the layer (see Figure~\ref{fig2}, panel E and F).  For convenience, the sign of the source terms for each set is shown in columns six and seven of Table~\ref{tab1}.
Since the sources $\hat{Q}_2$ and $\hat{Q}_{3,T}$ are divided by $\mathrm{Kn}_0$ in System~(\ref{sys8}), their absolute values increase with decreasing Knudsen number.


The number density always increases with the distance (see panel A in Figure~\ref{fig2}) and reaches its maximum value at the astropause (wall). The small increase of density by $\approx$ 10\% exists even in gas dynamical case ($\mathrm{Kn}_0 \rightarrow \infty$). This is weakly observed in Figure~\ref{fig2}, A due to the logarithmic scale. However, it is noticeable in
Figure~\ref{fig3} (panel A, black curve). The smaller the Knudsen number, the more significant the increase of density toward the astropause. The ratios of the number densities at the astropause (where the maximum is reached) and at the shock wave (where the minimum is reached) are $1.33$ for $\mathrm{Kn}_0 = 2.93$ (this value of Knudsen number corresponds to the heliosphere),  $1.66$ for $\mathrm{Kn}_0 = 1.63$;  $3.08$ for $\mathrm{Kn}_0 = 0.87$; and $8.41$ for $\mathrm{Kn}_0 = 0.6$. 

The positions of the shock wave (TS) are marked in Figure~\ref{fig2},~A. It can be concluded that the layer width increases as the Knudsen number decreases. The plasma velocity (panel B) decreases from the shock wave to the astropause almost linearly in all cases. 

In opposite to other parameters, the distribution of pressure (Figure~\ref{fig2}, C) is not always monotonic from the shock to the astropause. For high Knudsen numbers, the pressure increases toward the astropause. The smaller the Knudsen number, the smaller the pressure gradient. For $\mathrm{Kn}_0 = 0.6$ and $0.87$ the gradient of pressure changes sign and the pressure maximum moves away from the astropause. The temperature (panel D) decreases with decreasing Knudsen number.


To analyze the results above, we performed additional tests, in which we artificially turned off either the momentum source or thermal source (see Appendix~\ref{A1}). The results of the analyses can be summarized as follows (see Appendix~\ref{A1} for details):

1. The momentum source affects mainly the pressure distribution in the shock layer. In the gas dynamics case, the pressure increases towards the astropause, providing the deceleration of the flow. In the presence of atoms, a negative source of momentum leads to deceleration of the flow as well. Additional deceleration is not needed in this case and, therefore, the pressure gradient decreases (so that the total deceleration remains approximately the same). If the momentum source is strong enough, the pressure gradient changes the sign and may slightly accelerate the flow. The negative momentum source has almost no effect on the number density distribution, but significantly increases the shock layer thickness. 

2. The thermal source, on the contrary, has almost no effect on the pressure distribution but leads to a considerable decrease in the temperature and increase in the number density towards the astropause, and slightly reduces the thickness of the shock layer. 

The range of $[0.6 - 3]$ for Knudsen numbers that has been performed is not quite narrow as it may be thought. Indeed, the results of $\mathrm{Kn}_0 \approx 3$ show that the effect of charge exchange is quite small already. For larger $\mathrm{Kn}_0$ the solution is close to gas dynamics. For Knudsen numbers smaller than 0.6 the effect of neutrals becomes very strong. Our simplified approach is not valid anymore, and a self-consistent model should be employed. The applicability of the presented analysis to the heliosphere and other astrospheres is discussed in Section~\ref{relev}.



\subsection{Set 2 results}
\label{ss2}
\label{s2r}
\begin{figure*}
	\begin{minipage}[h!]{1.00\linewidth}
		\center{\includegraphics[width=1.0\linewidth]{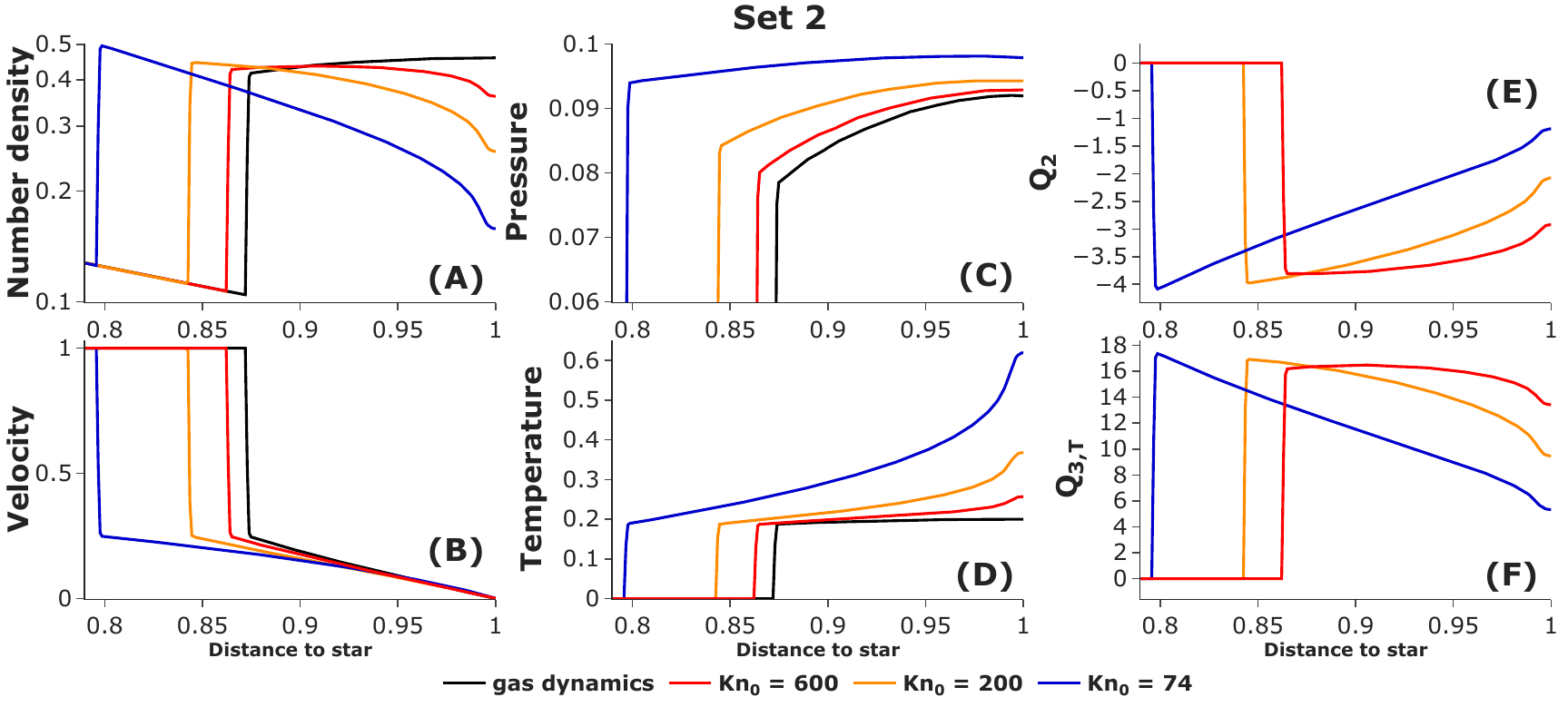} }
	\end{minipage}
	\caption{Distribution of dimensionless plasma parameters in the shock layer on the axis of symmetry (X-axis) for various values of the Knudsen number (including the gas dynamics case) for the distribution of hydrogen in the shock layer given in Table~\ref{tab1} (Set 2)
 }
	\label{fig3}
\end{figure*}

Figure~\ref{fig3} shows the distribution of plasma parameters along the axis of symmetry in the shock layer for the hydrogen parameters of Set 2 (see Table~\ref{tab1}). The atoms of this set are faster and hotter as compared with the atoms of Set 1. Comparing the source terms $Q_2$ and $Q_{3, T}$ With those of Set 1 (see, panels E and F in Figures~\ref{fig2} and \ref{fig3}) shows that the absolute value of $Q_2$ is in about 40 times larger for Set 2. The absolute values of $Q_{3, T}$ are about 200-250 times larger. As it follows from previous consideration, for $Kn_0 \approx 0.1-10$ we can not obtain any meaningful solution with such large source terms. However, in a dimensionless solution the right parts depend on the ratios of $\hat{Q}_2$, $\hat{Q}_{3, T}$, and Knudsen number $Kn_0$. Increasing the Knudsen number by a factor of 200 would reduce the source terms to reasonable values. 
Therefore, for Set 2 we vary the Knudsen number in the range of [$\sim$100,$\sim$600]. Due to hot temperature $\hat{T}_H$ chosen for Set 2 the thermal source $\hat{Q}_{3, T}$ is larger than the momentum source $\hat{Q}_2$, then dividing the terms by the Knudsen numbers from the given range, we make the effect of momentum exchange is negligible for the Set 2 and the main purpose of this set is to demonstrate the effect of heating the shock layer.

The results of Set 2 demonstrate as previously that the larger the Knudsen number, the smaller effects of charge exchange. The main effects can be summarized as follows:

1. Figure~\ref{fig3}, A  shows that the plasma number density decreases towards the astropause with the largest decrease for the runs with small $\mathrm{Kn}_0$. The minimum number density is reached at the astropause. The maximum is reached either inside the layer (see $\mathrm{Kn}_0 = 600$) or on the shock wave ($\mathrm{Kn}_0 = 74\ \mathrm{or}\ 200$). The number density maximum and minimum ratios are $1.21,\ 1.79,\ \mathrm{and}\ 3.14$ for $\mathrm{Kn}_0 = 600,\ 200, \ \mathrm{and}\  74$, respectively.

2. The layer width increases and the pressure gradient (Figure~\ref{fig3}, C) decreases with decreasing Knudsen number.

3. The velocity profile (panel B) deviates slightly from linear, and the temperature (Panel D) rises towards the astropause.

 

\subsection{Set 3 results}

\begin{figure*}
	\begin{minipage}[h!]{1.00\linewidth}
		\center{\includegraphics[width=1.0\linewidth]{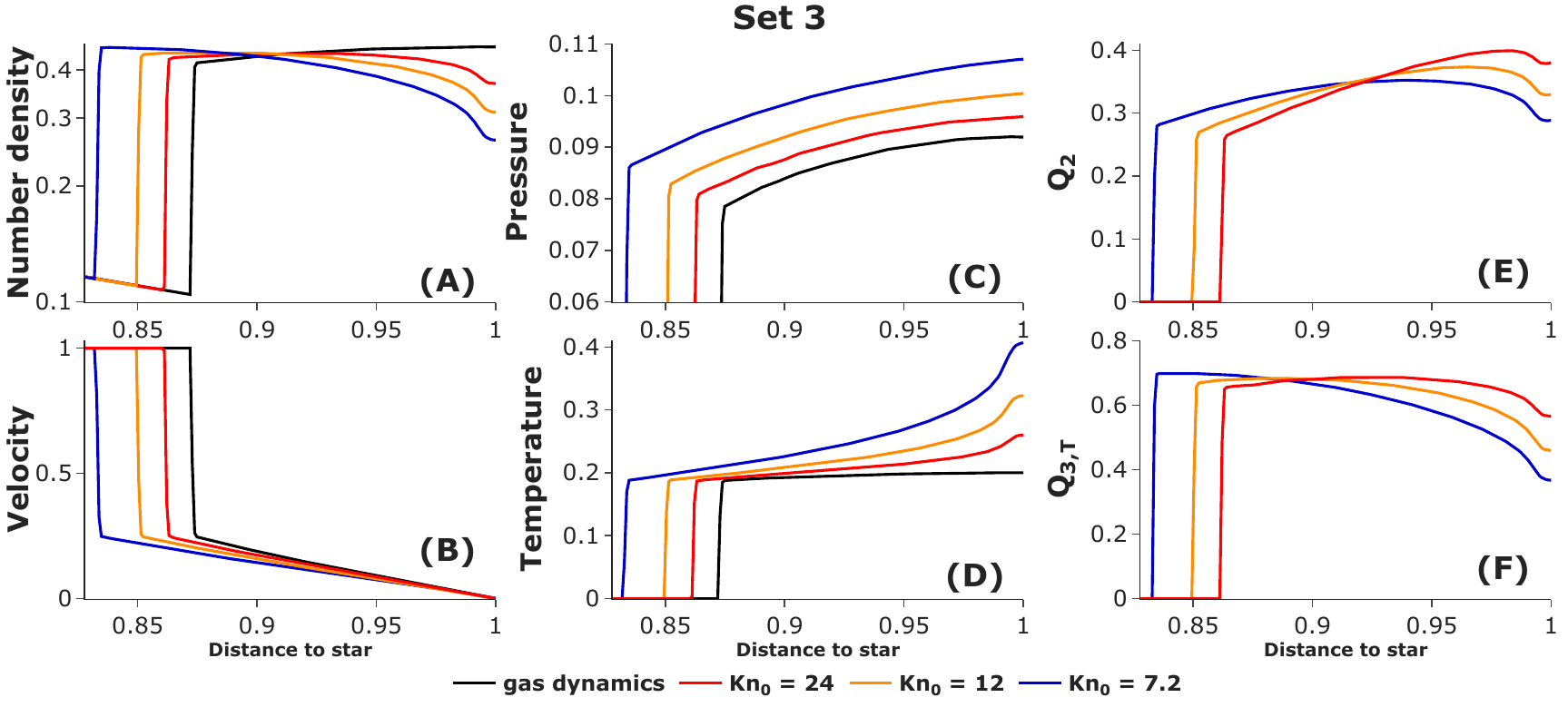} }
	\end{minipage}
	\caption{Distribution of dimensionless plasma parameters in the shock layer on the axis of symmetry (X-axis) for various values of the Knudsen number (including the gas dynamics case) for the distribution of hydrogen in the shock layer given in Table~\ref{tab1} (Set 3)
 }
	\label{fig4}
\end{figure*}

Figure~\ref{fig4} shows the distribution of plasma parameters along the axis of symmetry in the shock layer for the third set of hydrogen parameters (see Table~\ref{tab1}). In this case, we explore the combined effect of three populations of atoms (see Table~\ref{tab1}). The choice of such neutral parameters is not random and is related to the distributions of hydrogen populations in the heliosphere (see Section~\ref{relev}). The combined effect of these populations results in the positive momentum (panel E in Figure~\ref{fig4}) and thermal  (panel F) sources. 

In this case, the distributions of the density (panel A), velocity (panel B), and temperature (panel D) are quite similar to those of Set 2.
The main difference from Set 2 is the pressure distribution (panel C). It increases towards the astropause, providing stronger braking of the plasma by the pressure gradient. This behavior fully confirms the conclusions made above.

\section{Relevance to the heliosphere and particular astrospheres}
\label{relev}

\begin{table}
\centering
\caption{Mass loss rate ($\dot{M}$), characteristic distance to the astropause ($\mathrm{L_{HP}}$, calculated from the global model), Knudsen number ($\mathrm{Kn_0}$) for various stars. The terminal wind speed ($V_0$) and mean number density value in the inner heliosheath ($n_\mathrm{H}$) are considered to be 400 km/s and 0.105 $\mathrm{cm}^{-3}$, respectively. All data taken from \protect\cite{2005Wood, 2021Wood}}
\label{tab2}
\begin{tabular}{lccc}
\hline
Star&  $\dot{M}$  ($\dot{M}_\odot$)  & $\mathrm{L_{HP}}$ (AU)& $\mathrm{Kn_0}$ \\
\hline
\hline
Sun & 1  & 120 & 2.93 \\
\hline
ksi Boo & 5  & 260 & 1.42  \\
\hline
GJ 15AB & 10  & 400 & 0.93 \\
\hline
GJ 644AB & <5  & 125 & 2.96  \\
\hline
d Eri & 4  & 180 & 2.06  \\
\hline
YZ CMi & 30  & 880 & 0.42 \\
\hline
70 Oph & 100  & 1000 & 0.37  \\
\hline
GJ 338AB & 0.5  & 90 & 4.12 \\
\hline
\end{tabular}
\end{table}


The performed parametric study aims to explore the effects of charge exchange on the SL plasma flows in general.  However, it appears that the obtained results have immediate implications for 1) the most explored astrosphere which is our heliosphere, and 2) some other observable astrospheres.

\subsection{Heliosphere}

Let us start with the heliosphere. The run of Set 1 with $\mathrm{Kn}_0 = 2.93$ is relevant to the inner heliosheath (i.e. region 2 in Figure~\ref{fig1}, A). Indeed, assuming $V_0$~=~400 km/s and $L= 120$ AU that is relevant to the solar wind speed and the inner heliosheath thickness, we obtain dimensional parameters $V_\mathrm{H}$~=~-20.8 km/s and $T_\mathrm{H}$~=~14370 K. These values are close to the values of interstellar atoms (mixture of primary and secondary, see \cite{Izmodenov_2000}) at the inner heliosheath.

Our results (see Subsection~\ref{s1r}) show that the moderate increase of proton number density in the inner heliosheath observed by Voyager 2 at the distances between $\sim 94$ AU to $\sim 117$ AU may be caused by the negative thermal source (cooling) of plasma by charge exchange with interstellar atoms of hydrogen. The density increase can be seen in Figure 2 of \citep{Richardson2022} and should not be confused with a much larger increase in the so-called  Plasma Boundary Region (PBR), the physical nature of which is still unknown (see, section 5.2 of \citep{Richardson2022}).

Sets 2 and 3 have relevance to the outer heliosheath (the region 3 in Figure~\ref{fig1}, A).
Assuming $V_0$ = 26.4 km/s, in Set 2 we obtain dimensional parameters $V_\mathrm{H}$ = 73 km/s and $T_\mathrm{H}$ = 470000 K. These values correspond to hot atoms originating in the inner heliosheath (population 2). Such atoms are detected by IBEX as a globally distributed flux of energetic neutral atoms (ENAs, \cite{2009McComas}). 

In this case, our toy model shows that the plasma number density should increase beyond the heliopause, which has a direct connection with the heating of the plasma by atoms of population 2. It is interesting to note that the behavior of the plasma density in the shock layer obtained for Set 2 (see Subsection~\ref{s2r}) is consistent with the behavior of the interstellar plasma beyond the heliopause (i.e. in the outer heliosheath). For example, in the model of \cite{Alexashov_Izmodenov2005} (Figure~3), where the charge exchange was taken into account and magnetic fields were also neglected, the ratio of the number density maximum to its value at the astropause is 1.4 and the distance between corresponding locations is 25 AU. The Voyager 1 data shows this ratio of about 2.3, and the distance between the corresponding locations is about 17 AU (\cite{Richardson2022}, Figure~36). This may be considered evidence of a physical process that heats the plasma in the outer heliosheath near the heliopause. Moreover, this process must be more effective in the vicinity of the heliopause. Otherwise, the maximum density would be further from the heliopause. Therefore, the analogy between our model simulations and the Voyager 1 may serve as tentative proof that the heating of the interstellar plasma in the vicinity of the heliopause by the heliospheric ENAs.

The drop in plasma number density beyond the heliopause was discussed in \cite{Baranov_Malama_1993, Fuselier2013, Pogorelov2017, Pogorelov2021}. Nowadays, four different explanations of the plasma depletion layer (PDL) can be found in the literature: 
(1) the effect of the interstellar magnetic field which piled up near the heliopause, (2) the effect of charge exchange, (3) the gas-dynamic effect of the flow around a blunt body, (4) non-stationary effects. 
We argue that such a density drop does not exist in the gas-dynamic flows around blunt bodies (see, for example, \cite{Lyubimov1970}, volume 2, Figure 14.27).
Global models, which include all processes listed above \citep{izmodenov_alexashov2015, Izmodenov_Aleksashov_2020}, show that the depletion of the interstellar plasma number density toward the heliopause exists. However, the density gradient obtained in the models is much smaller than observed. This may be an indication that an additional heating process should play a role in the immediate vicinity of the heliopause.

The shock layer in the case of Set 3 is also relevant to the outer heliosheath. However, it represents the more realistic situation when the SL plasma flow experiences the combined effect of several populations. We chose parameters of these populations in a such way that they represent neutral solar wind (population 1), heliospheric ENAs (population 2), and primary interstellar atoms (population 4). Assuming the value of $V_0$ = 26.4 km/s, we obtain $V_\mathrm{H,1}$ = 322 km/s, $V_\mathrm{H,2}$ = 73 km/s, $V_\mathrm{H,3}$ = 26.4 km/s, and $T_\mathrm{H,1}$ = 145000 K, $T_\mathrm{H,2}$ =470000 K, and $T_\mathrm{H,2}$= 7450 K, respectively. The relative number densities also roughly correspond to the average number densities of the respective populations in the outer heliosheath ($10^{-4}\ \mathrm{cm}^{-3},\ 8\cdot 10^{-4}\ \mathrm{cm}^{-3},\ 0.07$ cm$^{-3}$). In this case, the primary interstellar atoms form a positive momentum source, making the pressure gradient in the outer shock layer significantly greater than for Set 2, which is in better agreement with global models of the heliosphere (see, e.g., \cite{izmodenov_alexashov2015}, Figure~4, B, blue dash-dot curve). However, they are not cold enough, so the heat source remains positive and the number density profile remains similar to the Set 2 results.

\subsection{Astrospheres}

The results of model calculations for Set 1 (see Subsection~\ref{s1r}) with the range of Knudsen numbers of [0.6 - 3.0] may also be relevant to various astrospheres. We analyze parameters of seven stars for which astrospheric Lyman-alpha absorption has been observed \citep{2005Wood, 2021Wood} and found that estimated values of the Knudsen numbers for the stars are also in the considered range (or close to the range, see Table~\ref{tab2}). Thus, with hydrogen number density of $0.106\ \mathrm{cm}^{-3}$ and star terminal velocity of $400\ \mathrm{km/s}$, the presented range describes astrospheres with sizes from $\sim$120 to $\sim$600 AU. If we take into account the generally accepted estimate for the size of the astrosphere: $L \approx \sqrt{\dfrac{\dot{M} V_0}{4 \pi \rho_{ISM} V_{ISM}}}$ (see \cite{1990Baranov}), then stellar mass losses vary from $\sim$1 to $\sim$25 solar one. For stars of lower $\dot{M}$, the influence of neutrals is practically insignificant and the solution is close to the limiting gas-dynamic flow, which was also considered in this work.

The effect of plasma cooling, which leads to an increase in the number density and decrease in the temperature of the plasma in the inner shock layer toward the astropause that was discussed in Subsection~\ref{s1r} (for Set~1) has been also obtained 
for the global model for the YZ CMi star in \cite{2021Wood} (see Figure~8, panels~a~and~b).

It is also worthwhile to note, that the effect of plasma cooling during the charge exchange that we discussed for Set 1, may be quite similar for other types of cooling. For example, an increase in plasma number density near the astropause can be observed if radiative cooling in the outer shock layers of astrospherics is taken into account (see, for example, a parametric study of \cite{2021_Titova}, Figure~4a). For some specific astrospheres, the number density increase can be very large and localized in a very narrow layer near the astropause (see \cite{2022Light}, Figures~8,~10).

The density reduction toward the astropause due to heating was obtained for Set 2 (Subsection~\ref{s2r}). This effect was also observed for the global model of the GJ 15AB star for the outer shock layer (\cite{2021Wood}, Figure~5). Unfortunately, the plasma number density distributions are not presented there, but the maximum hydrogen number density is achieved near the bow shock. For such a large astrosphere (3.3 times larger than the heliosphere), this should mean that the maximum plasma number density is also much closer to the bow shock than to the astropause.

We hope that the effect of plasma heating will be observed in the outer astrosheaths of stars having large astrospheres, where the charge exchange of plasma with energetic neutral atoms is very effective, in the future.

\section{Summary}
\label{C}


In this paper, we performed parametric study and explored how the charge exchange of protons with atomic hydrogen changes the plasma flow in the shock layer. The results of the work can be summarized as follows:
\begin{itemize}
  \item The momentum source primarily influences the pressure distribution in the shock layer. In the gas dynamics case, the pressure increases towards the astropause, providing a negative gradient to slow down the flow. In the presence of atomic hydrogen, the pressure at the astropause may increase even more if the momentum source is directed towards the astropause (accelerates the flow). On the contrary, in the case of the negative momentum source (decelerates the flow), the maximum pressure appears inside the shock layer or even at the shock wave.
  \item The momentum source also changes the width of the shock layer, a negative source increases the size of the layer, while a positive source decreases it.
  \item Heating of the shock layer leads to its expansion and a decrease in the number density of the plasma toward the astropause. Cooling, on the contrary, reduces the layer and increases the number density at the astropause. The heating/cooling source does not directly affect the pressure and velocity distributions. However, small changes in the distributions may be observed due to changes in the layer width.
\end{itemize}

The paper also discussed the application of the presented results to the heliosphere and various astrospheres (see Section~\ref{relev}). In particular, in the case of the heliosphere, the discussed effects of charge exchange can partially explain the increase in plasma number density towards the heliopause in the inner heliosheath, as well as the decrease in plasma number density near the heliopause in the outer heliosheath (PDL). For the outer shock layers of large astrospheres, we predict a decrease in plasma number density near the astropause and its increase near the bow shock due to the heating of the layer by astrospheric ENAs; we believe that such observational data will appear in the future.

\section*{Acknowledgements}
Korolkov Sergey thanks the Theoretical Physics and Mathematics Advancement Foundation "BASIS" grant 22-8-3-42-1.

The research is carried out using the equipment of the shared research facilities of HPC computing resources at Lomonosov Moscow State University \citep{Voevod2019}.

 \section{Data Availability}
 \label{D}
The data underlying this article will be shared on reasonable request to the corresponding author.



\bibliographystyle{mnras}
\bibliography{example} 



\appendix

\section{Separation of sources}
\label{A1}

\begin{figure*}
	\begin{minipage}[h!]{1.00\linewidth}
		\center{\includegraphics[width=1.0\linewidth]{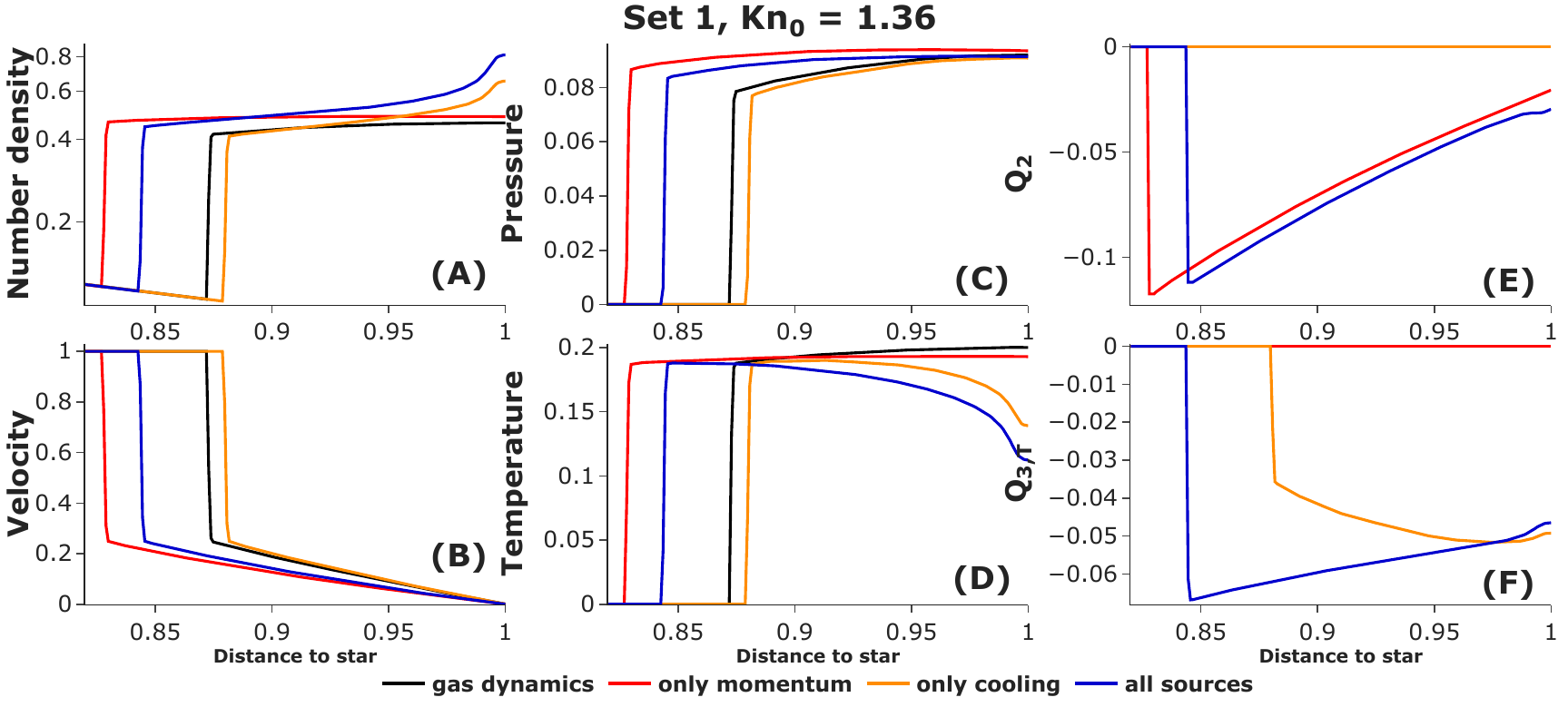} }
	\end{minipage}
	\caption{Distribution of dimensionless plasma parameters in the shock layer on the axis of symmetry (X-axis) for $\mathrm{Kn}_0$ = 1.36 for the distribution of hydrogen in the SL given in  Table~\ref{tab1} (Set 1) and the pure gas dynamics case (black curves). The red, orange, and blue curves correspond to only the source of momentum, energy, and both, respectively
 }
	\label{fig5}
\end{figure*}

\begin{figure*}
	\begin{minipage}[h!]{1.00\linewidth}
		\center{\includegraphics[width=0.8\linewidth]{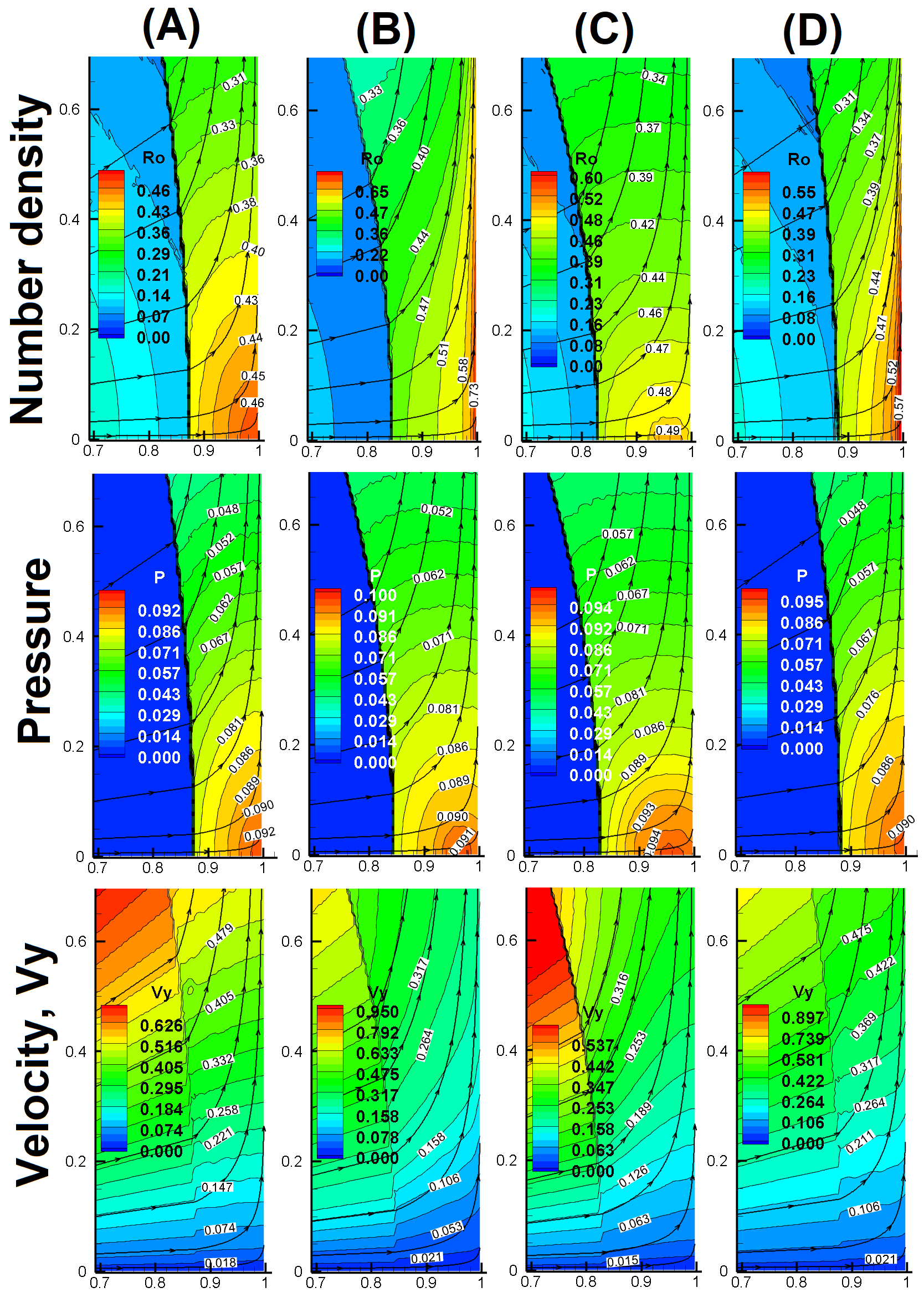} }
	\end{minipage}
	\caption{Streamlines and isolines of number density, pressure, and perpendicular velocity components for calculation from Figure~\ref{fig5}. Columns A - D correspond to pure gas dynamics, all sources, only momentum and only cooling, respectively.}
	\label{fig6}
\end{figure*}

In this Appendix, we present the results of two additional test calculations performed with the parameters of Set 1 (see Table~\ref{tab1}). In the first (1) test-run, we turn off the thermal source $Q_{3, T}$ while the momentum source $Q_2$ is calculated as usual (see, Equation~\ref{yr2}). In the second (2) test-run, we, in opposite, turn off the momentum source $Q_{2}$ while the thermal source $Q_{3, T}$ is calculated as usually (see, Equation~\ref{yr3}).
 Such tests help us to separate the effects of the momentum and thermal sources on the plasma flow in the shock layer.

Figure~\ref{fig5} shows the results of the test-runs for $\mathrm{Kn}_0 = 1.36$. For the sake of comparison, we also plot (3) results of the run in the absence of atoms, when $Q_2=0$ and $Q_{3, T} = 0$ (black curves), and (4) results of the full-run when both momentum and thermal sources are taken into account (blue curves).


It is clearly seen from the comparison that the thermal source of $Q_{3, T}$ (which is negative in the considered case) is responsible for the decrease of the temperature toward the astropause. The pressure does not change significantly for test-run 2 (orange curves) as compared with the pure gas dynamic case (3). The increase of number density towards the astropause is explained by the behavior of pressure, temperature, and the state equation $p= 2 n k_B T$.  Therefore, we conclude that the thermal source is responsible for significant changes in the temperature and density distribution. Also, as it is seen from the Figure, the thickness of the shock layer was slightly reduced in test-run 2 (as compared with the pure gas dynamic case). This is explained by the reduction of thermal energy from the shock layer.

In opposite, the momentum source (test-run 2, red curves) does not affect temperature and density distribution. At the same time, it noticeably reduces the pressure gradient within the shock layer. To balance the reduction of pressure, the shock moves toward the star and, therefore, the thickness of the shock layers increases quite strongly.


Figure~\ref{fig6} shows two-dimensional distributions of number density, pressure, and velocity components perpendicular to the axis of symmetry. The results are presented for the pure gas dynamic case (column A), for both sources include (column B), for test-run 1 (column C), and test-run 2 (column D). 

From these plots, we pointed out one interesting observation. In the presence of the thermal source (columns B and D), the plasma number density has maximum at the astropause. In the case of test-run 1 (column C), the number density maximum slightly moved away from the astropause, but this maximum is much less pronounced than in columns B and D. Also, in the test-run 1 and for the full-run of Set 1 (column B), the pressure maximum is not at the astropause. 

The presented above distributions are stationary solutions that are the result of balance between different terms in mass, momentum, and energy equations. To better understand how the stationary balance is established let us consider how pure gas-dynamic flow in the shock layer is transformed under the effects of the momentum and thermal sources.  
Let us start with the thermal source $Q_{3,T} < 0$ (i.e. cooling). 
The cooling leads to a decrease in temperature in the shock layer as gas moves from the shock to astropause. The temperature decrease results in a reduction of pressure toward the astropause, and, therefore, the negative pressure gradient along the X-axis is reduced (becomes smaller in absolute terms). The smaller the pressure gradient, the smaller the deceleration of plasma in the shock layer.

At the same time, the pressure gradient along the astropause (Y-axis) also decreases and the plasma evacuates from the stagnation point region slower than in the gas-dynamic case. As a result, the plasma number density increases toward the astropause (in the upwind) and, also, along the astropause (Y-axis). The increase in number density leads to an increase of the pressure, so a new balanced distribution of pressure is established close to those of the gas-dynamic case. The width of the shock layer is reduced slightly due to the reduction of thermal energy. As a result, a new balanced solution is established, and it is shown in Figure~\ref{fig6}, D.  

A similar scenario can be described for the momentum source. The negative momentum source slows down the plasma flow, so the negative pressure gradient is not needed to decelerate the flow to the astropause. This increases the pressure in the shock layer, which pushes the shock wave closer to the star to achieve a new balanced solution at the shock wave.


\bsp	
\label{lastpage}
\end{document}